\documentclass[journal=jpccck,manuscript=article]{achemso}
\usepackage[utf8]{inputenc}

\usepackage{chemformula}
\usepackage[T1]{fontenc}
\usepackage{amsmath}
\usepackage{graphicx}
\usepackage[labelfont=bf]{caption}
\usepackage{comment}
\usepackage{indentfirst}

\author{Tao Wu}
\affiliation[East China University of Science and Technology]
{Department of Mechanical and Power Engineering, East China University of Science and Technology, Shanghai 200237, China}
\altaffiliation{These authors contributed equally.}

\author{Bo Liu}
\email{boliu@ecust.edu.cn}
\affiliation[East China University of Science and Technology]
{Department of Mechanical and Power Engineering, East China University of Science and Technology, Shanghai 200237, China}
\altaffiliation{These authors contributed equally.}

\author{Haohao Hao}
\affiliation[Southern University of Science and Technology]
{Multicomponent Fluids Group, Center for Complex Flows and Soft Matter Research \& Department of Mechanics and Aerospace Engineering, Southern University of Science and Technology, Shenzhen 518055, Guangdong, China}

\author{Xuehua Zhang}
\email{xuehua.zhang@ualberta.ca}
\affiliation[University of Alberta]
{Department of Chemical \& Materials Engineering, University of Alberta, Edmonton, Alberta T6G 1H9, Canada}

\author{Fang Yuan}
\affiliation[East China University of Science and Technology]
{Department of Mechanical and Power Engineering, East China University of Science and Technology, Shanghai 200237, China}

\author{Huanshu Tan}
\email{tanhs@sustech.edu.cn}
\affiliation[Southern University of Science and Technology]
{Multicomponent Fluids Group, Center for Complex Flows and Soft Matter Research \& Department of Mechanics and Aerospace Engineering, Southern University of Science and Technology, Shenzhen 518055, Guangdong, China}

\author{Qiang Yang}
\email{qyang@ecust.edu.cn}
\affiliation[East China University of Science and Technology]
{Department of Mechanical and Power Engineering, East China University of Science and Technology, Shanghai 200237, China}

\title[Cascade coalescence sustains bubble retention]
{Cascade coalescence dynamically sustains bubble retention near gas-evolving surfaces}

\abbreviations{VOF, volume of fluid}

\begin{document}

\begin{abstract}

Bubble detachment from solid surfaces governs heat, mass, and charge transport across technologies vital to clean energy, including high-current-density water electrolysis and boiling thermal management. At high gas fluxes, however, bubbles remain trapped at active surfaces despite immense buoyancy, severely restricting mass transfer and increasing energy losses. Here, we show that this unexpected surface retention originates from cascade coalescence between unequal-sized bubbles. High-speed observations around microelectrodes demonstrate that when a rising bubble merges with a smaller surface-attached successor, its trajectory abruptly reverses, accelerating toward the substrate at nearly two orders of magnitude above its rising speed. Direct numerical simulations and scaling analysis reveal that asymmetric interfacial retraction during merging generates non-canceling viscous stresses, producing a net downward impulse toward the smaller bubble. Repeated cascade coalescence events accumulate these transient impulses into a steady, time-averaged retaining force capable of opposing buoyancy three to four orders of magnitude beyond quasistatic limits. Our findings establish bubble coalescence as a previously unrecognized mechanism that dynamically sustains bubble retention under high gas flux.

\end{abstract}

\section{Introduction}

Bubble growth and departure from solid surfaces underpin critical clean energy processes, ranging from green hydrogen generation during high-current-density water electrolysis~\cite{ZhangChemRev2024,KemplerChemRev2024,DengEScience2025,LeeJoule2024} to boiling-based thermal management for high-power microelectronics~\cite{DhirAnnuRevFluidMech1998,ZhangIJHMT2022,InanluSciAdv2024}.
In these systems, the rapid removal of gas or vapor bubbles from active surfaces is paramount. Persistent bubble accumulation shields active sites, restricts mass transfer, and drastically increases electrolysis overpotentials and thermal resistance.
Under quasistatic regimes, bubble retention is dictated by the classical force balance between buoyancy and capillary forces along the three-phase contact line.
For a circular contact line of diameter $d_{\mathrm{c}}$, the pinning capillary force scales as $F_{\mathrm{cap}} = \pi d_{\mathrm{c}} \gamma \sin\theta$, yielding the Fritz departure criterion~\cite{FritzPhysZ1935,VogtElectrochimActa1989} which predicts detachment at tens of micrometres on highly wetting substrates~\cite{ZhaoACSAMI2024,IwataJoule2021,DarbandRSER2019}.
At high reaction rates, however, bubbles remain tightly bound to the electrode, growing to millimetric scales despite immense buoyancy forces~\cite{WuEScience2025,BashkatovJACS2024,WuIJHE2025}.
Because buoyancy scales with volume ($F_{\mathrm{b}} \propto D^3$), a size increase from $50~\mu\text{m}$ to $1~\text{mm}$ amplifies the detaching force by a factor of $8{,}000$.
Conventional models cannot explain what retains these bubbles against forces three to four orders of magnitude beyond the classical limit, revealing a fundamental gap in our understanding of high-flux bubble retention mechanisms.

Several secondary forces can influence bubble detachment, yet none account for this strong surface retention under high gas fluxes.
In boiling systems, rapid evaporation generates growth-induced hydrodynamic forces~\cite{ZengIJHMT1993,KlausnerIJHMT1993} and localized vapor recoil stresses that drive dry-spot expansion~\cite{NikolayevPRL2006,ZhangNatCommun2023}.
However, these thermal mechanisms are largely absent during electrochemical gas evolution, where gas production is limited by electrochemical reaction rates and bubble growth is far slower~\cite{ZhangChemRev2024,KemplerChemRev2024}.
Interfacial charging and localized concentration or temperature gradients can also induce electrostatic and Marangoni stresses~\cite{ParkNatChem2023,BashkatovPRL2019,LuJPowerSources2024,MeulenbroekElectrochimActa2024,MassingElectrochimActa2019,HossainElectrochimActa2020}.
While these field forces contribute under specific regimes, estimates based on established force models show that, under the present conditions, they remain at least an order of magnitude below the buoyancy of the retained millimetric bubbles~\cite{LuCEJ2024,ParkNatChem2023,MeulenbroekElectrochimActa2024,MassingElectrochimActa2019,HossainElectrochimActa2020,HossainPRE2022} (Supplementary Note~10).
More importantly, none of these single-bubble mechanisms explain why surface retention intensifies with bubble number density and coalescence frequency~\cite{WuEScience2025,BashkatovJACS2024}.
This intensification implies that coalescence itself may be related to the  mechanism sustaining bubble retention.

However, bubble coalescence, which is ubiquitous across gas-evolving systems, is classically understood to promote bubble detachment rather than surface retention.~\cite{ZhangCellRepPhysSci2024,IwataLangmuir2022,LvPRL2021}. When two bubbles merge, the excess interfacial energy released during coalescence is converted into kinetic energy, which can propel the merged bubble away from the substrate~\cite{IwataLangmuir2022,LvPRL2021,ZhangPRL2025}. Recent electrochemical experiments have revealed an apparently opposite outcome: a departing bubble can coalesce with a smaller, surface-attached successor, after which the merged bubble moves back toward the electrode in a ``comeback'' mode.~\cite{WuEScience2025,BashkatovJACS2024}
Despite these opposite migration directions, both coalescence-induced detachment and comeback have been interpreted using the same capillary--inertial framework~\cite{BashkatovJACS2024,IwataLangmuir2022,LvPRL2021}. In this framework, viscous traction is generally neglected because of the low Ohnesorge number ($Oh=\mathcal{O}(10^{-2})$)~\cite{EggersAnnuRevFluidMech2025,EggersJFM1999,ThoroddsenPhysFluids2005}. The merged bubble is expected to move toward the volume-weighted center of its parent bubbles, as quantitatively verified for two bubbles growing on a surface~\cite{LvPRL2021,ChenAIChEJ2017}. For the comeback configuration, this center is closer to the electrode surface, predicting a downward displacement of the merged bubble~\cite{BashkatovJACS2024,LvPRL2021}.
However, the apparent retaining force inferred from the downward displacement scales as $D^{-2}$ and rapidly diminishes relative to buoyancy. The capillary-inertial framework therefore predicts that coalescence alone cannot sustain bubble retention~\cite{BashkatovJACS2024}.

Here we show how coalescence-induced return develops into sustained bubble retention. High-speed imaging around gas-evolving microelectrodes reveals that a rising bubble reverses its motion after coalescing with a smaller surface-attached bubble, accelerating toward the surface at nearly two orders of magnitude above its rising speed.
Direct numerical simulations and scaling analysis show that, although liquid inertia governs the rapid neck expansion, the viscous stresses generated by unequal interfacial retraction do not cancel.
Their time integral produces a finite impulse directed toward the smaller bubble.
Repeated coalescence events convert these transient impulses into a time-averaged retaining force capable of opposing the buoyancy of bubbles far beyond the quasistatic departure limit.
Consistently, suppressing the supply of surface smaller bubbles causes retained bubbles to escape for a few milliseconds.
These results identify cascade coalescence as mechanism that dynamically sustains bubble retention under high gas flux.

\clearpage
\section{Results and discussion}

\subsection{Coalescence reverses bubble motion}

\begin{figure}[htbp]
\centering
\includegraphics[width=\linewidth]{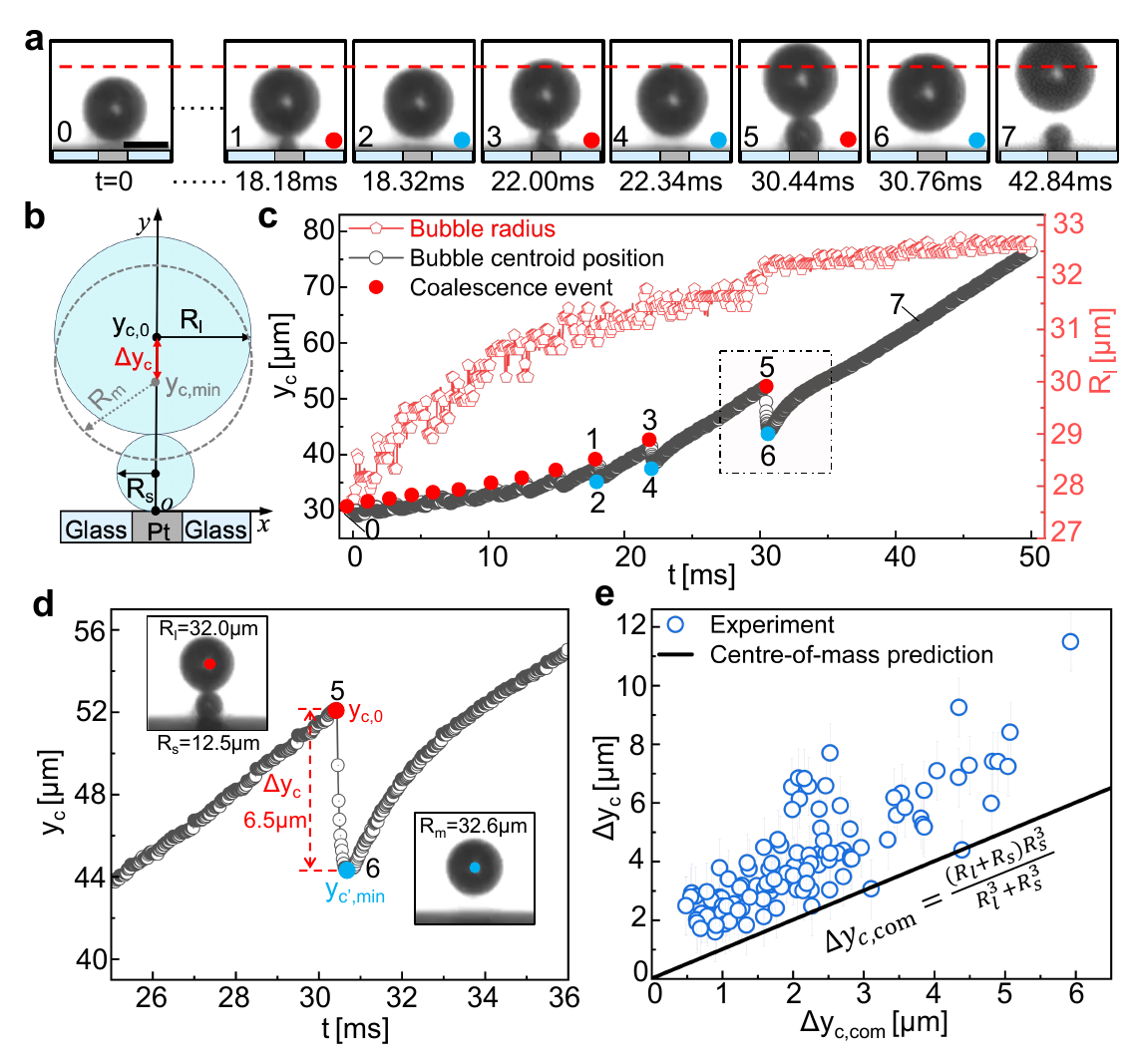}
\caption{
\textbf{Coalescence reverses bubble motion.}
\textbf{(a)} Time-resolved snapshots of a hydrogen bubble undergoing repeated coalescence with surface microbubbles after detachment. Each coalescence event drives the bubble towards the electrode before final departure. Frames 1--7 show the last three drag-back events; $T$ denotes the elapsed time after the initial detachment.
\textbf{(b)} Geometric definitions. $R_l$, $R_s$, and $R_m$ denote the radii of the detached bubble, surface bubble, and merged bubble, respectively. The $x$ and $y$ coordinates are parallel and normal to the electrode surface; $y_{c,0}$ and $y_c'$ denote centroid positions immediately before and after coalescence.
\textbf{(c)} Bubble centroid position $y_c$ (black) and radius $R_l$ (red) during cascade coalescence, showing repeated downward displacement and stepwise growth.
\textbf{(d)} Final coalescence event in (c), showing a drag-back distance $\Delta y_c\approx6.5~\mu\mathrm{m}$ within $0.3~\mathrm{ms}$.
\textbf{(e)} Measured $\Delta y_c$ compared with the center-of-mass prediction\cite{LvPRL2021}, $\Delta y_{c,\mathrm{com}}$,over $R_s=7.2-14.9~\mu\mathrm{m}$ and $R_l=22.7-32~\mu\mathrm{m}$.
}
\label{fig:fig1}
\end{figure}

We generated hydrogen bubbles on a $50~\mu\mathrm{m}$-diameter Pt microelectrode operated at a low current of $|I|=0.01$--$0.015~\mathrm{mA}$ in an electrolyte containing $0.5~\mathrm{M}$ \ch{H2SO4} and $0.1~\mathrm{M}$ \ch{Na2SO4}, producing a sparse population of surface microbubbles. Under these conditions, individual coalescence events between a detached bubble and a single growing surface microbubble were resolved using high-speed microscopy (Supplementary Note~1). The low-current condition minimized thermal and solutal\cite{ParkNatChem2023,MassingElectrochimActa2019,HossainElectrochimActa2020}, as well as growth-induced\cite{ZengIJHMT1993,NikolayevPRL2006}, contributions to bubble motion (Supplementary Note~2). The subsequent trajectory therefore provides a direct measure of the coalescence-induced bubble motion.

A representative sequence in Fig.~\ref{fig:fig1}a illustrates the evolution of a detached hydrogen bubble after leaving the electrode. Instead of escaping into the bulk, the detached bubble repeatedly encounters and coalesces with newly formed surface microbubbles. Remarkably, each coalescence event reverses the bubble trajectory, driving the merged bubble back towards the electrode (Supplementary Video~1). This repeated return delays bubble escape and enables subsequent coalescence with newly formed surface bubbles, sustaining a cascade of coalescence events.

To quantify this motion, we define the coordinate system and geometric parameters shown in Fig.~\ref{fig:fig1}b. The electrode surface is taken as $y=0$, with the positive $y$ direction pointing away from the electrode. The quantities $R_s$ and $R_l$ denote the radii of the smaller surface-growing bubble and the larger detached bubble immediately before coalescence, respectively, while $R_m$ denotes the radius of the merged bubble. Throughout the analysis, we tracked the centroid position $y_c(t)$ of the evolving bubble during the coalescence cascade. Before each coalescence event, $y_c(t)$ represents the detached bubble; after coalescence, it represents the newly merged bubble.

The evolution of $y_c(t)$ is shown in Fig.~\ref{fig:fig1}c. Each coalescence event appears as a sharp downward excursion in $y_c(t)$, accompanied by a discrete increase in bubble size. Starting from the initial detachment at $t=0$, the bubble undergoes more than ten coalescence-induced returns before finally escaping at $t=30.76~\mathrm{ms}$. Over this coalescence cascade, its radius increases from $27.5$ to approximately $32.6~\mu\mathrm{m}$, corresponding to a 67\% increase in volume. Thus, this coalescence cascade sustains a dynamic near-surface retention state, where repeated coalescence events simultaneously drive bubble growth and maintain its proximity to the electrode.

We then examine whether the observed drag-back motion can be explained by the conventional capillary--inertial picture, which has successfully described bubble neck expansion, interfacial retraction, and jumping dynamics after coalescence\cite{LvPRL2021,WeonPRL2012}. In this picture, viscous effects are neglected, and momentum conservation therefore predicts that the merged bubble relocates towards the volume-weighted center\cite{LvPRL2021},
\begin{equation}
\Delta y_{c,\mathrm{com}}
=
\frac{(R_l+R_s)R_s^3}{R_l^3+R_s^3},
\label{eq:center-of-mass}
\end{equation}
of the two parent bubbles immediately before coalescence (Supplementary Note~3), as validated for neighboring surface-growing bubbles\cite{LvPRL2021,ChenPRE2020,WeonPRL2012}. To test this prediction, we quantify the experimentally observed drag-back distance for each event as $\Delta y_c=y_{c,0}-y_{c,\min}$, where $y_{c,0}$ is the centroid position of the detached bubble immediately before coalescence and $y_{c,\min}$ is the turning point corresponding to the maximum downward displacement of the merged bubble (Fig.~\ref{fig:fig1}b,d).

However, our experiments show that the capillary--inertial prediction substantially underestimates the observed drag-back distance. For the representative event shown in Fig.~\ref{fig:fig1}d, with $R_l=32.0~\mu\mathrm{m}$ and $R_s=12.5~\mu\mathrm{m}$, the merged bubble travels approximately $6.5~\mu\mathrm{m}$ towards the electrode within $0.3~\mathrm{ms}$, whereas the predicted relocation distance $\Delta y_{c,\mathrm{com}}$ is only $2.5~\mu\mathrm{m}$. This substantial discrepancy persists across the explored size ratios, with the measured drag-back distances (blue circles) consistently exceeding the predicted values (black solid line) by factors of two to three (Fig.~\ref{fig:fig1}e). This deviation from the prediction suggests an additional wall-directed contribution generated during coalescence.
\clearpage
\subsection{Coalescence imparts an impulsive downward velocity}

\begin{figure}[htbp]
\centering
\includegraphics[width=\linewidth]{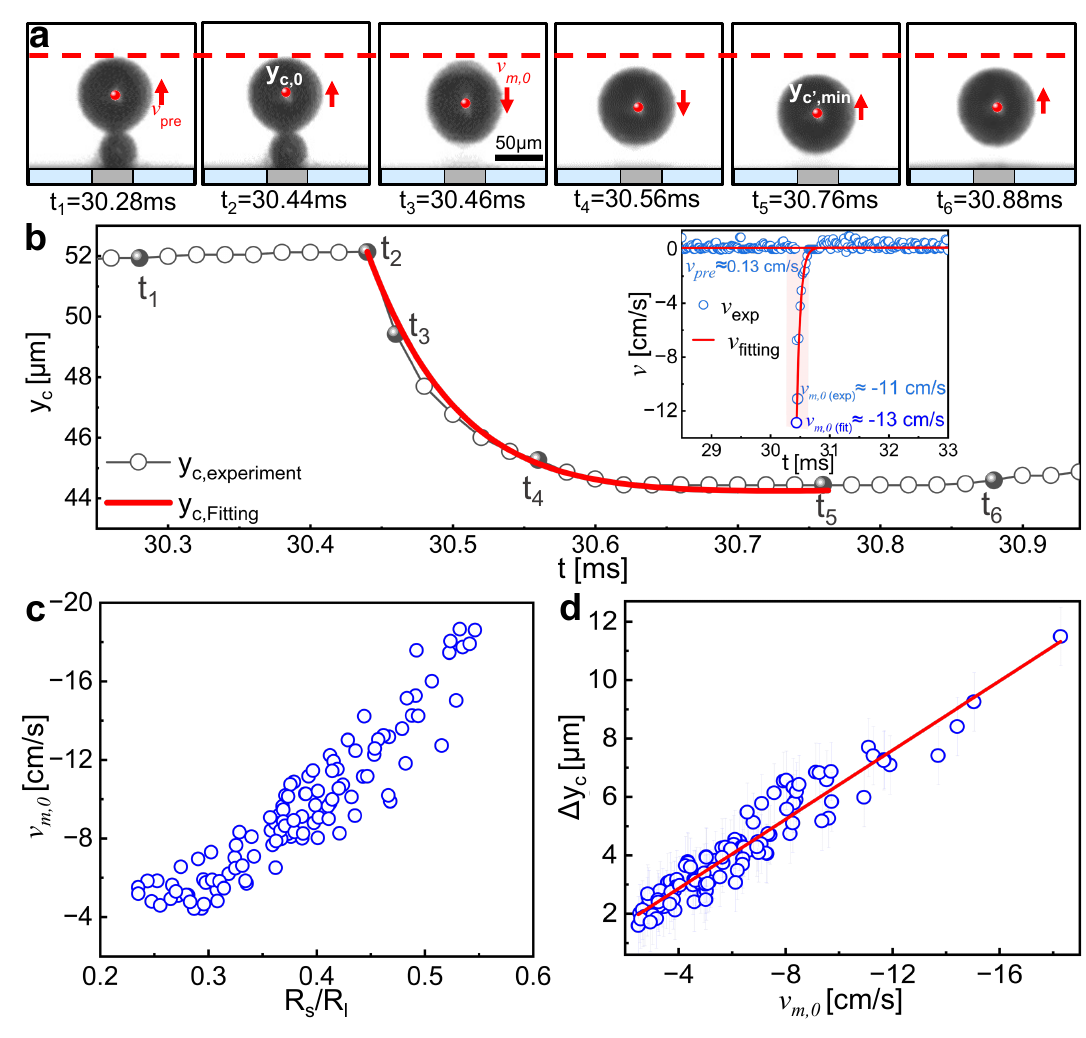}
\caption{
\textbf{Coalescence reverses bubble motion through an impulsive downward velocity.}
\textbf{(a)} High-speed image sequence of an axial coalescence event between a detached bubble and a surface microbubble. The detached bubble rises before coalescence with velocity $v_{\mathrm{pre}}$, while the merged bubble acquires an initial downward velocity $v_{m,0}$ after coalescence. Red arrows indicate the bubble-motion direction; the red dashed line marks the pre-coalescence reference height.
\textbf{(b)} Time-resolved centroid trajectory corresponding to (a). Symbols denote experimental measurements, and the solid red line shows a fitted decelerating trajectory used to determine $v_{m,0}$. The inset gives the initial downward velocity extracted by trajectory fitting  ($\sim13~\mathrm{cm\,s^{-1}}$ for the representative event).
\textbf{(c)} Initial downward velocity $v_{m,0}$ as a function of the parent-bubble size ratio $R_s/R_l$.
\textbf{(d)} Drag-back distance $\Delta y_c$ as a function of $v_{m,0}$.
}
\label{fig:fig2}
\end{figure}

 We next ask how coalescence produces such a large drag-back distance. Beyond the predicted relocation, our experiments reveal an unexpected downward velocity after coalescence. In the representative event shown in Fig.~\ref{fig:fig2}a, the detached bubble is rising with velocity $v_{\mathrm{pre}}\approx0.13~\mathrm{cm\,s^{-1}}$ before coalescence (Supplementary Note~4), while the surface-growing bubble remains attached to the electrode and continues to grow. Within the capillary--inertial framework, part of the upward momentum of the rising parent bubble should persist through coalescence, causing the resulting bubble to continue moving upward immediately afterward. Contrary to this prediction, high-speed imaging reveals a pronounced velocity reversal: upon coalescence, the merged bubble acquires a large downward velocity and moves towards the electrode. This downward motion increases the drag-back distance until the bubble reaches a turning point and subsequently resumes buoyancy-driven ascent (Fig.~\ref{fig:fig2}b and Supplementary Video~2).

We determine this initial downward velocity by fitting the post-coalescence trajectory with a force-balance model containing buoyancy, bulk viscous drag, and near-wall lubrication resistance (Eq.~\eqref{eq:trajectory} and Supplementary Note~5):
\begin{equation}
 m_{\mathrm{eff,m}}v_m\frac{\mathrm{d}v_m}{\mathrm{d}y_c}
 =
 F_b-
 \left(
 4\pi\mu R_m+
 \frac{4\pi\mu R_m^2}{H}
 \right)v_m.
\label{eq:trajectory}
\end{equation}
Here, upward motion is taken as positive, $v_m(t)=\mathrm{d}y_c/\mathrm{d}t$, and $v_{m,0}=v_m(0^+)$ denotes the merged-bubble velocity immediately after coalescence. The effective inertia of the merged bubble is represented by its added mass\cite{MagnaudetAnnuRevFluidMech2000}, $m_{\mathrm{eff,m}}=C_a\rho_l(4/3)\pi R_m^3$, where $C_a=1/2$ is the added-mass coefficient of a spherical bubble and $\rho_l$ is the liquid density\cite{MagnaudetAnnuRevFluidMech2000}. The quantities $F_b$, $\mu$, $R_m$, and $H=y_c-R_m$ denote the buoyant force, liquid viscosity, merged-bubble radius, and instantaneous bubble--electrode gap, respectively. Without introducing any additional retaining force after coalescence, this model reproduces the observed downward deceleration, turning point, and subsequent rise, suggesting that the motion is a deceleration process from an initially imposed velocity. For the representative event, the model gives $v_{m,0}\approx-13~\mathrm{cm\,s^{-1}}$, approximately 100 times the magnitude of the pre-coalescence velocity (Supplementary Note~4).

Over the explored ranges $R_l\approx30$--$50~\mu\mathrm{m}$ and $R_s\approx5$--$20~\mu\mathrm{m}$, the magnitude of the downward initial velocity spans approximately $4$--$20~\mathrm{cm\,s^{-1}}$ and increases systematically with $R_s/R_l$ (Fig.~\ref{fig:fig2}c). The drag-back distance increases approximately linearly with $|v_{m,0}|$, consistent with the post-coalescence deceleration model (Fig.~\ref{fig:fig2}d and Supplementary Note~5). Therefore, the large drag-back displacement is determined by the unexpectedly large downward velocity acquired during the short coalescence event.
\newpage
\subsection{Asymmetric viscous impulse reverses bubble motion}

\begin{figure}[htbp]
\centering
\includegraphics[width=\linewidth]{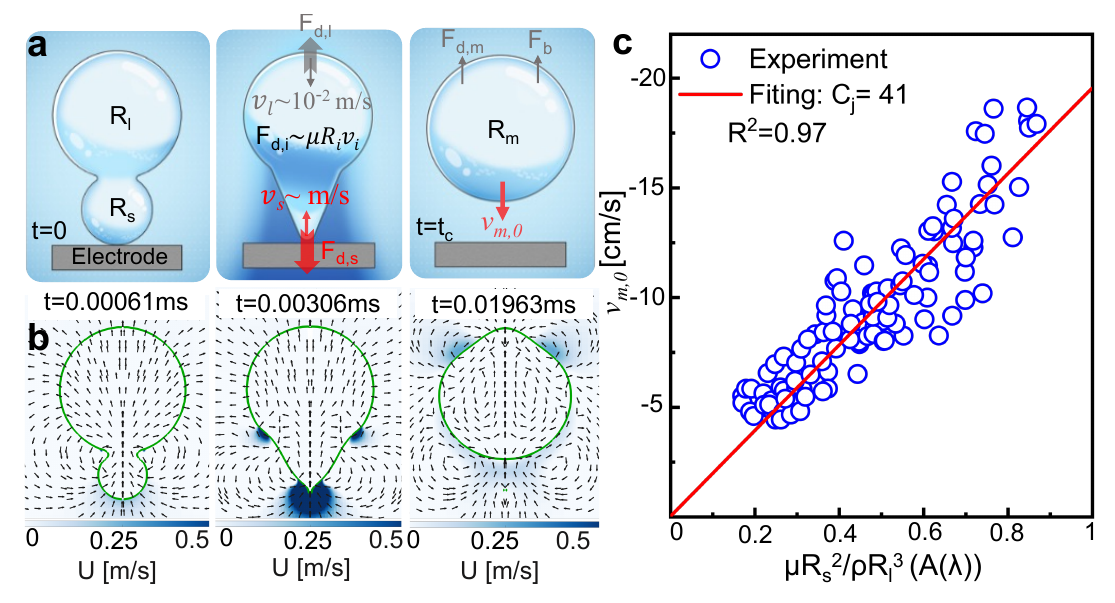}
\caption{
\textbf{An asymmetric viscous impulse generates the post-coalescence downward velocity.}
\textbf{(a)} Schematic illustration of the asymmetric viscous-impulse mechanism during axial coalescence of unequal-sized bubbles. Unequal retraction dynamics generate an imbalance in viscous contributions, producing a residual wall-directed impulse $J_v$ that gives rise to the initial downward velocity $v_{m,0}$ of the merged bubble.
\textbf{(b)} Direct numerical simulations of the coalescence-induced flow, showing localized strong interfacial motion near the smaller bubble.
\textbf{(c)} Collapse of the measured $v_{m,0}$ using the viscous-impulse scaling; $\rho$ on the horizontal axis denotes the liquid density $\rho_l$.
}
\label{fig:fig3}
\end{figure}

We propose that the observed rapid downward motion is driven by a net viscous impulse generated by size-asymmetric coalescence (Fig.~\ref{fig:fig3}a). Following initial bridge formation, the resolved interfacial retraction is predominantly capillary--inertial, consistent with the small Ohnesorge numbers ($\mathrm{Oh}i=0.021$--$0.044$; Supplementary Note~6)\cite{EggersAnnuRevFluidMech2025,EggersJFM1999}. 
The characteristic retraction velocity associated with parent bubble $i$ therefore scales as $v_i\sim[\sigma/(\rho_lR_i)]^{1/2}$\cite{EggersAnnuRevFluidMech2025,EggersJFM1999}, yielding $v_l/v_s\sim(R_s/R_l)^{1/2}=\lambda^{1/2}$, where $\lambda=R_s/R_l$. In a frame translating with the merged bubble, the effective retraction displacement ratio follows $l_{d,l}/l_{d,s}\sim\lambda^3$ (Supplementary Note~7). Thus, the smaller surface bubble undergoes faster and more extensive interfacial retraction than the larger detached bubble, producing strongly asymmetric retraction dynamics (Fig.~\ref{fig:fig3}b and Supplementary Note~8).

Although viscous stresses are subdominant in the instantaneous force balance, their time-integrated contributions need not cancel, generating a net directional contribution. The viscous force associated with interface of bubble $i$ scales as $F_{d,i}\sim\mu R_iv_i$ and acts over $t_i\sim l_{d,i}/v_i$, yielding an impulse $J_{v,i}\sim \mu R_i l_{d,i}$. 
Taking the direction toward the electrode as positive, the net viscous impulse is
\begin{equation}
J_v
\int\left(F_{d,s}-F_{d,l}\right)\mathrm{d}t
\approx
C_j\mu R_s^2A(\lambda),
\label{eq:viscous-impulse}
\end{equation}
where $A(\lambda)=1-\lambda^2$ quantifies the retraction asymmetry and $C_j$ is a dimensionless prefactor accounting for interfacial geometry, temporal evolution, and wall effects. 
Because the smaller bubble lies on the electrode side and its dominant retraction is directed away from the electrode, the corresponding viscous reaction is electrode-directed and exceeds the opposing contribution by a factor $\lambda^{-2}$.
The size-asymmetric impulse vanishes as $\lambda\rightarrow1$.  Thus, capillary--inertial dynamics govern the rapid shape evolution, whereas the non-cancelling viscous impulse biases the residual motion toward the electrode.

The net viscous impulse acts on the effective inertia of the merged bubble, which is dominated by the added mass of the surrounding liquid, setting its initial velocity after coalescence, $J_v=m_{\mathrm{eff},m}v_{m,0}$. For $R_s\ll R_l$, $R_m\simeq R_l$, yielding
\begin{equation}
v_{m,0}
=
-\frac{3C_j}{2\pi}
\frac{\mu}{\rho_l}
\frac{R_s^2}{R_l^3}
A(\lambda).
\label{eq:initial-velocity}
\end{equation}
The model predicts both the wall-directed sign of $v_{m,0}$ and its dependence on parent-bubble size asymmetry, with the detailed derivation provided in Supplementary Note~7. The measured $v_{m,0}$ values collapse onto a linear relation when plotted against $(\mu/\rho_l)(R_s^2/R_l^3)A(\lambda)$ (Fig.~\ref{fig:fig3}c), with the fitted slope giving $C_j=41$. 
This agreement supports the proposed asymmetric viscous-impulse mechanism and provides the prefactor used below to quantify the retaining force generated by repeated coalescence.

\subsection{Repeated coalescence sustains a retaining force}
\begin{figure}[htbp]
\centering
\includegraphics[width=\linewidth]{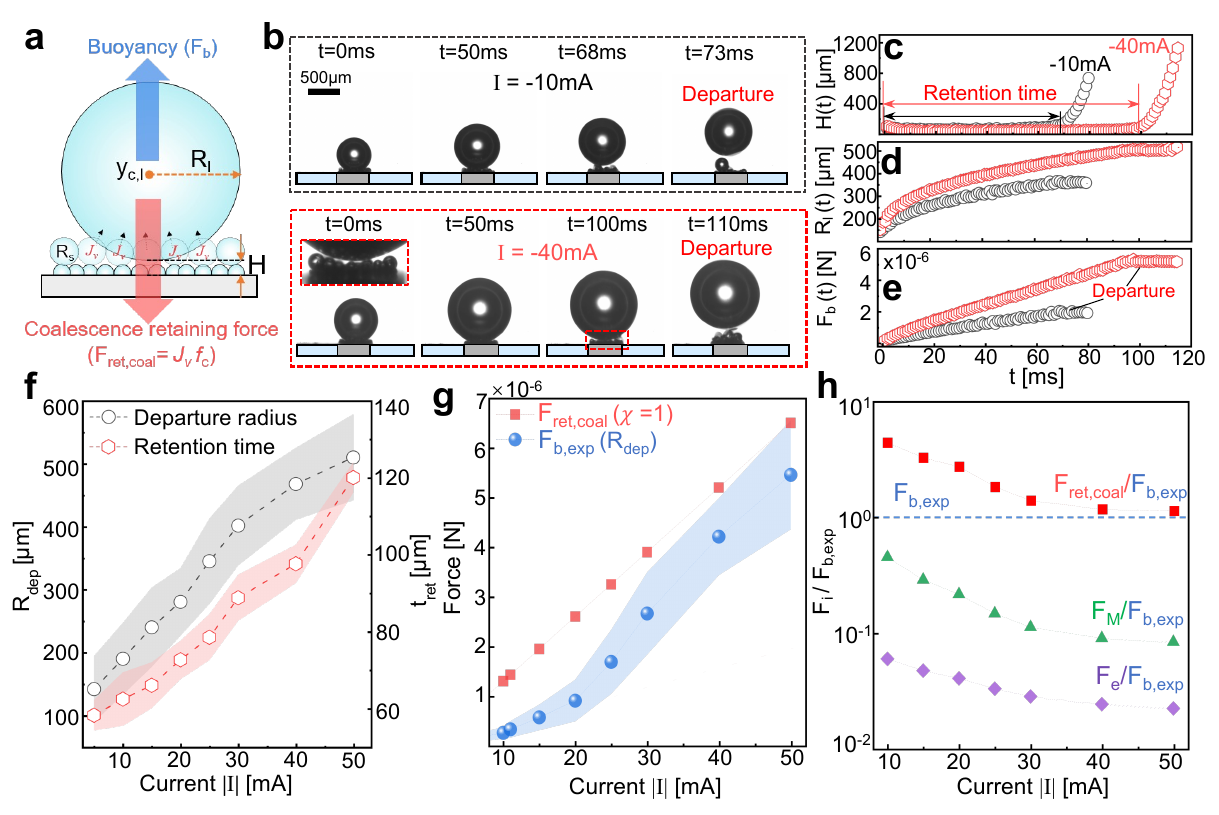}
\caption{
\textbf{Repeated surface-microbubble coalescence sustains bubble retention.}
\textbf{(a)} Schematic of a detached bubble retained above the surface-microbubble carpet on a $500~\mu\mathrm{m}$-diameter Pt microelectrode. Repeated coalescence supplies a downward impulse flux, producing a sustained retaining force $F_{\mathrm{ret,coal}}$ that opposes buoyancy $F_b$. The bubble--electrode separation is defined as $H=y_{c,l}-R_l$.
\textbf{(b)} Representative high-speed sequences showing bubble retention and final departure at $-10$ and $-40~\mathrm{mA}$. Inset: enlarged view of the surface-microbubble carpet.
\textbf{(c--e)} Temporal evolution of $H(t)$, bubble radius $R_l(t)$, and buoyancy $F_b(t)$. The plateau in $H(t)$ marks the retained state; the abrupt rise marks threshold crossing and escape.
\textbf{(f)} Departure radius $R_{\mathrm{dep}}$ and retention time $t_{\mathrm{ret}}$ as functions of current magnitude $|I|$.
\textbf{(g)} Estimated coalescence-induced retaining force $F_{\mathrm{ret,coal}}$ at the upper-bound limit $\chi=1$, compared with the experimental buoyancy $F_{b,\mathrm{exp}}$ at departure.
\textbf{(h)} Comparison of $F_{\mathrm{ret,coal}}$, Marangoni force $F_M$, and electrostatic force $F_e$, all normalized by $F_{b,\mathrm{exp}}$.
}
\label{fig:fig4}
\end{figure}

Having identified the wall-directed impulse generated by a single coalescence event, we now return to the high-flux gas-generation regime, where a detached large bubble resides above a continuously replenished carpet of surface microbubbles (Fig.~\ref{fig:fig4}a and Supplementary Video~3). In this regime, coalescence is no longer an isolated event. Instead, repeated coalescence events occur at a frequency $f_c$, converting discrete impulses into a sustained retaining force through time-averaged impulse transfer. Relating $f_c$ to the gas-generation rate through Faraday's law\cite{Faraday1839}, $f_c=\chi(3|I|R_gT_g)/(8\pi FP_gR_s^3)$, gives (Supplementary Note~9)
\begin{equation}
F_{\mathrm{ret,coal}}
=
J_vf_c
\approx
\chi\frac{3C_j\mu|I|R_gT_g}{8\pi FP_gR_s}A(\lambda).
\label{eq:retaining-force}
\end{equation}
Here, $F_{\mathrm{ret,coal}}$ denotes the equivalent sustained retaining force generated by repeated coalescence, and $\chi$ represents the effective impulse-transfer efficiency. The ideal limit $\chi=1$ corresponds to complete participation of generated microbubbles and perfect alignment of the resulting impulses towards the electrode, whereas incomplete participation or imperfect alignment gives $\chi<1$. The remaining quantities are defined above, with $I$ denoting the applied current, $R_g$ the universal gas constant, $T_g$ the gas temperature, $F$ the Faraday constant, and $P_g$ the internal bubble pressure. For $R_s\ll R_l$, $A(\lambda)\rightarrow1$, indicating that $F_{\mathrm{ret,coal}}$ is governed primarily by the gas-generation rate and characteristic microbubble size, with weak dependence on the retained-bubble radius. In contrast, buoyancy grows as $R_l^3$, implying that the retained bubble remains near the electrode until its buoyancy exceeds the coalescence-induced retaining force.

We tested this prediction using a $500~\mu\mathrm{m}$-diameter Pt microelectrode. Representative high-speed sequences at $-10$ and $-40~\mathrm{mA}$ show that a detached bubble remains suspended above the surface-microbubble carpet rather than escaping immediately into the bulk (Fig.~\ref{fig:fig4}b). The retained state is quantified by tracking the bubble--electrode separation, $H=y_{c,l}-R_l$, where $y_{c,l}$ and $R_l$ are the centroid position and radius of the retained large bubble, respectively. As shown in Fig.~\ref{fig:fig4}c, $H(t)$ remains nearly constant during retention and increases abruptly upon departure, defining the transition from retention to escape (Supplementary Videos~4 and 5). Meanwhile, repeated coalescence continuously increases $R_l(t)$ and the corresponding buoyancy $F_b(t)$ (Fig.~\ref{fig:fig4}d,e). The bubble is therefore dynamically retained by ongoing coalescence until its increasing buoyancy exceeds the coalescence-induced retaining force. Increasing $|I|$ from $5$ to $50~\mathrm{mA}$ extends the retention time $t_{\mathrm{ret}}$ from $58$ to $120~\mathrm{ms}$ and increases the departure radius from $140$ to $510~\mu\mathrm{m}$ (Fig.~\ref{fig:fig4}f), demonstrating that stronger gas generation sustains retention against larger buoyant loads.

We then compare the coalescence-induced retaining force with buoyancy at departure to test whether repeated coalescence can sustain retention. Using the independently fitted $C_j=41$ (Fig.~\ref{fig:fig3}c), a representative microbubble radius $R_s=10~\mu\mathrm{m}$, the measured departure radius $R_{\mathrm{dep}}$, and the ideal impulse-transfer limit $\chi=1$, the estimated $F_{\mathrm{ret,coal}}$ reaches the $10^{-6}~\mathrm{N}$ scale and slightly exceeds the experimental buoyancy at departure (Fig.~\ref{fig:fig4}g). This upper-bound estimate demonstrates that repeated coalescence can, in principle, provide sufficient retaining force under high gas flux. In comparison, even deliberately generous estimates of thermal and solutal Marangoni forces~\cite{ParkNatChem2023,MeulenbroekElectrochimActa2024,MassingElectrochimActa2019} and electrostatic forces~\cite{BashkatovPRL2019,LuJPowerSources2024} are about an order of magnitude smaller than buoyancy (Fig.~\ref{fig:fig4}h and Supplementary Note~10). This force hierarchy supports repeated surface-microbubble coalescence as the principal source of the retaining impulse flux when sustained gas generation maintains the microbubble source.

We further reproduce the coalescence-induced retention process in a pure-water microcapillary system, without electrochemical gas generation, applied electric fields, or reaction-induced thermal and solutal gradients (Supplementary Note~11 and Supplementary Video~7). In a near-horizontal orifice configuration, the detached bubble repeatedly coalesces with supplied microbubbles while remaining near the outlet. The vertical component of the coalescence-induced impulse counteracts buoyancy over successive events, until the bubble escapes when buoyancy exceeds the accumulated retaining effect. This observation demonstrates that coalescence-induced retention is not specific to electrochemical gas generation.

\subsection{Pulsed gas generation switches the impulse-flux balance}

\begin{figure}[htbp]
\centering
\includegraphics[width=\linewidth]{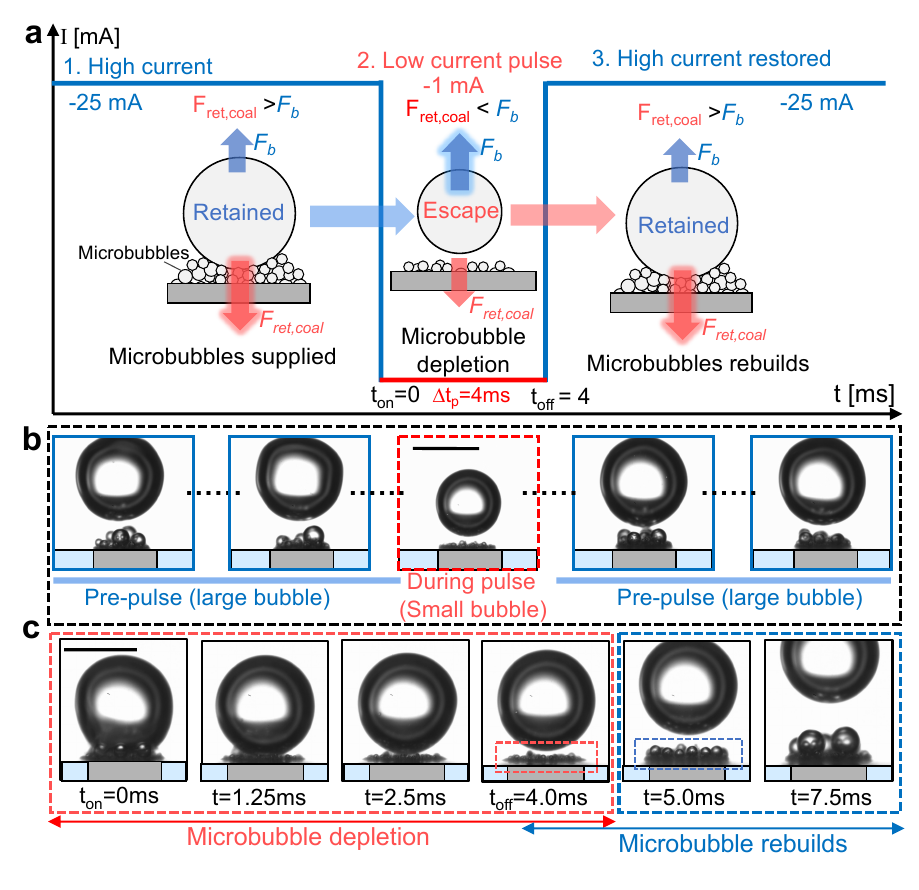}
\caption{
\textbf{A brief current pulse switches bubble retention by interrupting the surface-microbubble supply.}
\textbf{(a)} Schematic of the pulse-perturbation experiment. A bubble retained under high-current operation ($I_{\mathrm{high}}=-25~\mathrm{mA}$) is briefly exposed to a low-current pulse ($I_{\mathrm{low}}=-1~\mathrm{mA}$, $\Delta t_p=4~\mathrm{ms}$). The reduced gas-generation rate decreases the surface-microbubble supply and weakens the coalescence-induced retaining force.
\textbf{(b)} High-speed snapshots showing premature bubble departure at a smaller size after the low-current pulse compared with the surrounding high-current condition.
\textbf{(c)} Short-time image sequence resolving depletion and recovery of the surface-microbubble carpet. During the pulse ($t=0$--$4.0~\mathrm{ms}$), the carpet progressively disappears; after high-current operation is restored ($t=4.0$--$7.5~\mathrm{ms}$), the microbubble population is rapidly replenished.
}
\label{fig:fig5}
\end{figure}

We next explore the tunability of the coalescence-induced retaining force $F_{\mathrm{ret,coal}}$ by regulating the surface-microbubble source. A defining feature of this force is its dynamic origin: it is sustained only by repeated coalescence events that continuously deliver wall-directed impulses, so its magnitude depends on the availability of newly generated surface microbubbles. Briefly suppressing gas production should therefore deplete the surface-microbubble carpet, reduce the coalescence frequency $f_c$, and hence weaken $F_{\mathrm{ret,coal}}$. Once this force falls below buoyancy, the detached bubble should escape prematurely (Fig.~\ref{fig:fig5}a).

We tested this prediction on a $500~\mu\mathrm{m}$-diameter Pt microelectrode operated at $I_{\mathrm{high}}=-25~\mathrm{mA}$, where sustained gas generation at high current produced a dense surface-microbubble carpet and prolonged bubble retention. A brief low-current pulse ($I_{\mathrm{low}}=-1~\mathrm{mA}$ for $\Delta t_p=4~\mathrm{ms}$) reduced the gas-generation rate to one twenty-fifth of its high-current value, rapidly depleting the surface-microbubble population available for repeated coalescence. In the representative experiment shown in Fig.~\ref{fig:fig5}b, the retained bubble escapes shortly after the current is reduced, demonstrating that the coalescence-induced retaining force is no longer sufficient to balance buoyancy once the microbubble supply is interrupted. High-speed imaging directly visualizes the depletion of the surface-microbubble carpet beneath the retained bubble during the pulse (Fig.~\ref{fig:fig5}c and Supplementary Video~6). Quantitative tracking further confirms that the pulse reduces the retention time and departure radius (Supplementary Note~12). Thus, pulsed operation establishes the surface-microbubble supply as a direct control parameter for the coalescence-induced retaining force.
\clearpage
\section{Conclusions}

In summary, our work establishes cascade bubble coalescence, traditionally viewed as an ejection mechanism, as a dynamic hydrodynamic anchor that governs bubble retention under high gas fluxes.
By demonstrating that asymmetric interfacial retraction produces a sustained, time-averaged viscous impulse, these findings provides a new mechanism for multiphase control at solid-liquid interface: rather than relying  on static surface properties like contact angle or macro-morphology, bubble transport can be dynamically regulated by modulating local coalescence events through electrolyte rheology, targeted micro-texturing, and controlled nucleation dynamics.

Beyond bubble retention, repeated coalescence provides a persistent source of interfacial perturbations under high-gas-flux conditions. A millimeter-scale bubble may result from up to $10^5$ microbubble coalescence events({Supplementary Note 13}), each delivering a localized impulse that renews near-interface flows and modifies the local chemical environment. These high-frequency perturbations can enhance interfacial transport through repeated flow renewal, while also imposing high-cycle interfacial mechanical loading that may contribute to long-term material degradation, such as precious-metal loss from electrodes during water electrolysis. These insights should substantially reshape the understanding, modelling, and control of high-gas-flux interfaces across boiling, water electrolysis, catalytic gas evolution, and gas--liquid reactors.

\clearpage
\section{Experimental Methods}

\subsection{Electrolyte solution}
The electrolytes were prepared from \ch{H2SO4} (98\%, Suprapur, Merck), \ch{Na2SO4} (99\%, Suprapur, Merck), and Milli-Q water (resistivity $\geq18.2~\mathrm{M\Omega\,cm}$).

\subsection{Microelectrode experimental system}
Hydrogen bubbles were generated on Pt microelectrodes under galvanostatic conditions in a three-electrode electrochemical cell. A Hg/Hg$_2$SO$_4$ electrode was used as the reference electrode, and a Pt wire was used as the counter electrode. Two microelectrode sizes were used. The low-current single-coalescence experiments in Figs.~\ref{fig:fig1}--\ref{fig:fig3} were performed on a $50~\mu\mathrm{m}$-diameter Pt microelectrode, where sparse surface microbubbles enabled individual coalescence events to be resolved. The high-current retention and pulse experiments in Figs.~\ref{fig:fig4} and \ref{fig:fig5} were performed on a $500~\mu\mathrm{m}$-diameter Pt microelectrode, where dense surface-microbubble carpets produced sustained coalescence-induced bubble retention. The detailed microelectrode setup, electrode preparation, imaging configuration, and image-processing procedure are provided in Supplementary Note~1.

\subsection{Experimental data acquisition}
In the analysis, each individual coalescence event was treated as an independent data point. For every event, the radius of the newly detached parent bubble ($R_l$), the surface parent microbubble ($R_s$), the merged bubble ($R_m$), and the maximum centroid relocation distance ($\Delta y_c$) were quantified. Because parent-bubble sizes vary from event to event, this approach naturally samples a broad range of size ratios.

Owing to the temporal and spatial resolution limits of high-speed imaging, the analysis focuses on the final three coalescence events prior to bubble departure, for which the contours of the surface microbubbles can be reliably resolved. To ensure statistical robustness and measurement reliability, more than 200 independent coalescence events---corresponding to over 200 distinct parent-bubble size ratios---were analyzed and used for model fitting and scaling analysis. Error bars shown in the figures represent an image-based measurement uncertainty of $\pm2$ pixels ($\sim2.24~\mu\mathrm{m}$).

\subsection{Image analysis and data processing}
Image analysis and data processing were performed on the original high-speed image sequences. Bubble segmentation and size quantification were conducted using a deep-learning-based image-processing pipeline implemented in Python, based on the Cellpose framework\cite{PachitariuNatMethods2022,StringerNatMethods2021} (Supplementary Note~1).

\subsection{Numerical method}
Direct numerical simulations were performed using the open-source Basilisk framework to simulate the coalescence of two bubbles. A two-phase Navier--Stokes model was employed, and the gas--liquid interface was tracked using the volume-of-fluid (VOF) method (Supplementary Note~8).

\section{Data availability}
Source data underlying the figures are provided with this paper. Additional raw data, including the original high-speed imaging datasets, are available from the corresponding author upon reasonable request.

\section{Author contributions}
All authors wrote and read the paper. All authors conceived and designed the experiments. T.W.: Writing---original draft; Writing---review and editing; Methodology; Investigation. B.L.: Writing---review and editing; Conceptualization and model development; Funding acquisition; Supervision. H.H.: Methodology; Simulation; Writing---review and editing. X.Z.: Writing---review and editing. F.Y.: review and editing; Methodology. H.T.: Writing---review and editing; Conceptualization and model development; Funding acquisition; Simulation supervision. Q.Y.: Writing---review and editing; Funding acquisition; Supervision.

\section{Competing interests}
The authors declare that they have no competing interests.

\begin{acknowledgement}
We thank Prof. Qiang Sun from the ARC Centre of Excellence for Nanoscale BioPhotonics, RMIT University, and Prof. Bin Wang from East China University of Science and Technology for fruitful discussions.This work was financially supported by the National Natural Science Foundation of China (Grant Nos. 52025103, 22178099, and 12588301), the Shanghai Pilot Program for Basic Research (Grant No. 22TQ1400100-11), and the Guangdong Basic and Applied Basic Research Foundation (Grant No. 2024A1515010509 and 2024A1515010614).
\end{acknowledgement}

\begin{suppinfo}
\begin{itemize}
  \item Supplementary Information Note 1--13.
  \item Supplementary Videos 1--7.
\end{itemize}
\end{suppinfo}
\clearpage
\bibliography{ref}

@article{ZhangChemRev2024,
  author = {Zhang, Lenan and Iwata, Ryuichi and Lu, Zhengmao and Wang, Xuanjie and D{\'i}az-Mar{\'i}n, Carlos D. and Zhong, Yang},
  title = {Bridging Innovations of Phase Change Heat Transfer to Electrochemical Gas Evolution Reactions},
  journal = {Chemical Reviews},
  volume = {124},
  pages = {10052--10111},
  year = {2024}
}

@article{KemplerChemRev2024,
  author = {Kempler, P. A. and Coridan, R. H. and Luo, L.},
  title = {Gas Evolution in Water Electrolysis},
  journal = {Chemical Reviews},
  volume = {124},
  number = {19},
  pages = {10964--11007},
  year = {2024},
  doi = {10.1021/acs.chemrev.4c00211}
}

@article{DengEScience2025,
  author = {Deng, Lingao and Jin, Liming and Yang, Luyu and Feng, Chenchen and Tao, An and Jia, Xianlin and Geng, Zhen and Zhang, Cunman and Cui, Xiangzhi and Shi, Jianlin},
  title = {Bubble evolution dynamics in alkaline water electrolysis},
  journal = {eScience},
  volume = {5},
  pages = {100353},
  year = {2025}
}

@article{LeeJoule2024,
  author = {Lee, Jason K. and Babbe, Finn and Wang, Guanzhi and Tricker, Andrew W. and Mukundan, Rangachary and Weber, Adam Z. and Peng, Xiong},
  title = {Nanochannel electrodes facilitating interfacial transport for {PEM} water electrolysis},
  journal = {Joule},
  volume = {8},
  pages = {2357--2373},
  year = {2024}
}

@article{DhirAnnuRevFluidMech1998,
  author = {Dhir, V. K.},
  title = {Boiling Heat Transfer},
  journal = {Annual Review of Fluid Mechanics},
  volume = {30},
  pages = {365--401},
  year = {1998}
}

@article{ZhangIJHMT2022,
  author = {Zhang, L. and Gong, S. and Lu, Z. and Cheng, P. and Wang, E. N.},
  title = {Boiling crisis due to bubble interactions},
  journal = {International Journal of Heat and Mass Transfer},
  volume = {182},
  pages = {121904},
  year = {2022}
}

@article{InanluSciAdv2024,
  author = {Inanlu, Mohammad Jalal and Ganesan, Vishwanath and Upot, Nithin Vinod and Wang, Chi and Suo, Zan and Rabbi, Kazi Fazle and Kabirzadeh, Pouya and Bakhshi, Alireza and Fu, Wuchen and Thukral, Tarandeep Singh and Belosludtsev, Valentin and Li, Jiaqi and Miljkovic, Nenad},
  title = {Unveiling the fundamentals of flow boiling heat transfer enhancement on structured surfaces},
  journal = {Science Advances},
  volume = {10},
  pages = {eadp8632},
  year = {2024}
}

@article{FritzPhysZ1935,
  author = {Fritz, W.},
  title = {Maximum volume of vapor bubbles},
  journal = {Physikalische Zeitschrift},
  volume = {36},
  pages = {379--384},
  year = {1935}
}

@article{VogtElectrochimActa1989,
  author = {Vogt, H.},
  title = {The problem of the departure diameter of bubbles at gas-evolving electrodes},
  journal = {Electrochimica Acta},
  volume = {34},
  pages = {1429--1432},
  year = {1989}
}

@article{ZhaoACSAMI2024,
  author = {Zhao, P. and Gong, S. and Zhang, C. and Chen, S. and Cheng, P.},
  title = {Roles of Wettability and Wickability on Enhanced Hydrogen Evolution Reactions},
  journal = {ACS Applied Materials \& Interfaces},
  volume = {16},
  number = {21},
  pages = {27898--27907},
  year = {2024},
  doi = {10.1021/acsami.4c02428}
}

@article{IwataJoule2021,
  author = {Iwata, Ryuichi and Zhang, Lenan and Wilke, Kyle L. and Gong, Shuai and He, Mingfu and Gallant, Betar M. and Wang, Evelyn N.},
  title = {Bubble growth and departure modes on wettable/non-wettable porous foams in alkaline water splitting},
  journal = {Joule},
  volume = {5},
  pages = {887--900},
  year = {2021}
}

@article{DarbandRSER2019,
  author = {Darband, G. B. and Aliofkhazraei, M. and Shanmugam, S.},
  title = {Recent advances in methods and technologies for enhancing bubble detachment during electrochemical water splitting},
  journal = {Renewable and Sustainable Energy Reviews},
  volume = {114},
  pages = {109300},
  year = {2019}
}

@article{WuEScience2025,
  author = {Wu, Tao and Liu, Bo and Hao, Haohao and Yuan, Fang and Zhang, Yu and Tan, Huanshu and Yang, Qiang},
  title = {Coalescence-induced late departure of bubbles improves water electrolysis efficiency},
  journal = {eScience},
  pages = {100472},
  year = {2025},
  doi = {10.1016/j.esci.2025.100472}
}

@article{BashkatovJACS2024,
  author = {Bashkatov, Aleksandr and Park, Sunghak and Demirk{\i}r, {\c C}ayan and Wood, Jeffery A. and Koper, Marc T. M. and Lohse, Detlef and Krug, Dominik},
  title = {Performance Enhancement of Electrocatalytic Hydrogen Evolution through Coalescence-Induced Bubble Dynamics},
  journal = {Journal of the American Chemical Society},
  volume = {146},
  number = {14},
  pages = {10177--10186},
  year = {2024},
  doi = {10.1021/jacs.4c02018}
}

@article{WuIJHE2025,
  author = {Wu, Tao and Zhu, Zijian and Liu, Yuxi and Zhang, Wanyi and Yuan, Fang and Liu, Bo and Yang, Qiang},
  title = {Electrolyte-regulated bubble coalescence governs detachment size and electrolysis efficiency in hydrogen evolution},
  journal = {International Journal of Hydrogen Energy},
  volume = {175},
  pages = {151345},
  year = {2025}
}

@article{ZengIJHMT1993,
  author = {Zeng, L. Z. and Klausner, J. F. and Mei, R.},
  title = {A unified model for the prediction of bubble detachment diameters in boiling systems---I. Pool boiling},
  journal = {International Journal of Heat and Mass Transfer},
  volume = {36},
  pages = {2261--2270},
  year = {1993}
}

@article{KlausnerIJHMT1993,
  author = {Klausner, J. F. and Mei, R. and Bernhard, D. M. and Zeng, L. Z.},
  title = {Vapor bubble departure in forced convection boiling},
  journal = {International Journal of Heat and Mass Transfer},
  volume = {36},
  pages = {651--662},
  year = {1993}
}

@article{NikolayevPRL2006,
  author = {Nikolayev, V. S. and Chatain, D. and Garrabos, Y. and Beysens, D.},
  title = {Experimental Evidence of the Vapor Recoil Mechanism in the Boiling Crisis},
  journal = {Physical Review Letters},
  volume = {97},
  pages = {184503},
  year = {2006}
}

@article{ZhangNatCommun2023,
  author = {Zhang, Limiao and Wang, Chi and Su, Guanyu and Kossolapov, Artyom and Aguiar, Gustavo Matana and Seong, Jee Hyun and Chavagnat, Florian and Phillips, Bren and Rahman, Md Mahamudur and Bucci, Matteo},
  title = {A unifying criterion of the boiling crisis},
  journal = {Nature Communications},
  volume = {14},
  pages = {2321},
  year = {2023}
}

@article{ParkNatChem2023,
  author = {Park, Sunghak and Liu, Luhao and Demirk{\i}r, {\c C}ayan and van der Heijden, Onno and Lohse, Detlef and Krug, Dominik and Koper, Marc T. M.},
  title = {Solutal Marangoni effect determines bubble dynamics during electrocatalytic hydrogen evolution},
  journal = {Nature Chemistry},
  volume = {15},
  pages = {1532--1540},
  year = {2023},
  doi = {10.1038/s41557-023-01294-y}
}

@article{BashkatovPRL2019,
  author = {Bashkatov, A. and Hossain, S. S. and Yang, X. and Mutschke, G. and Eckert, K.},
  title = {Oscillating Hydrogen Bubbles at {Pt} Microelectrodes},
  journal = {Physical Review Letters},
  volume = {123},
  pages = {214503},
  year = {2019},
  doi = {10.1103/PhysRevLett.123.214503}
}

@article{LuJPowerSources2024,
  author = {Lu, X. and Yadav, D. and Ma, B. and Ma, L. and Jing, D.},
  title = {Rapid detachment of hydrogen bubbles for electrolytic water splitting driven by combined effects of Marangoni force and electrostatic repulsion},
  journal = {Journal of Power Sources},
  volume = {599},
  pages = {234217},
  year = {2024}
}

@article{MeulenbroekElectrochimActa2024,
  author = {Meulenbroek, A. M. and Deen, N. G. and Vreman, A. W.},
  title = {Marangoni forces on electrolytic bubbles on microelectrodes},
  journal = {Electrochimica Acta},
  volume = {497},
  pages = {144510},
  year = {2024}
}

@article{MassingElectrochimActa2019,
  author = {Massing, Julian and Mutschke, Gerd and Baczyzmalski, Dominik and Hossain, Syed Sahil and Yang, Xuegeng and Eckert, Kerstin and Cierpka, Christian},
  title = {Thermocapillary convection during hydrogen evolution at microelectrodes},
  journal = {Electrochimica Acta},
  volume = {297},
  pages = {929--940},
  year = {2019}
}

@article{HossainElectrochimActa2020,
  author = {Hossain, S. S. and Mutschke, G. and Bashkatov, A. and Eckert, K.},
  title = {The thermocapillary effect on gas bubbles growing on electrodes of different sizes},
  journal = {Electrochimica Acta},
  volume = {353},
  pages = {136461},
  year = {2020}
}

@article{ZhangCellRepPhysSci2024,
  author = {Zhang, Bo and Wang, Yechun and Feng, Yuyang and Zhen, Canghao and Liu, Miaomiao and Cao, Zhenshan and Zhao, Qiuyang and Guo, Liejin},
  title = {Coalescence and detachment of double bubbles on electrode surface in photoelectrochemical water splitting},
  journal = {Cell Reports Physical Science},
  volume = {5},
  pages = {101837},
  year = {2024},
  doi = {10.1016/j.xcrp.2024.101837}
}

@article{IwataLangmuir2022,
  author = {Iwata, Ryuichi and Zhang, Lenan and Lu, Zhengmao and Gong, Shuai and Du, Jianyi and Wang, Evelyn N.},
  title = {How Coalescing Bubbles Depart from a Wall},
  journal = {Langmuir},
  volume = {38},
  number = {14},
  pages = {4371--4377},
  year = {2022},
  doi = {10.1021/acs.langmuir.2c00118}
}

@article{LvPRL2021,
  author = {Lv, Pengyu and Pe{\~n}as, Pablo and Le The, Hai and Eijkel, Jan and van den Berg, Albert and Zhang, Xuehua and Lohse, Detlef},
  title = {Self-Propelled Detachment upon Coalescence of Surface Bubbles},
  journal = {Physical Review Letters},
  volume = {127},
  number = {23},
  pages = {235501},
  year = {2021},
  doi = {10.1103/PhysRevLett.127.235501}
}

@article{WeonPRL2012,
  author = {Weon, B. M. and Je, J. H.},
  title = {Coalescence preference depends on size inequality},
  journal = {Physical Review Letters},
  volume = {108},
  pages = {224501},
  year = {2012}
}

@article{ChenPRE2020,
  author = {Chen, R. and Yu, H. W. and Zeng, J. and Zhu, L.},
  title = {General power-law temporal scaling for unequal-size microbubble coalescence},
  journal = {Physical Review E},
  volume = {101},
  pages = {023106},
  year = {2020}
}

@article{MagnaudetAnnuRevFluidMech2000,
  author = {Magnaudet, J. and Eames, I.},
  title = {The motion of high-Reynolds-number bubbles in inhomogeneous flows},
  journal = {Annual Review of Fluid Mechanics},
  volume = {32},
  pages = {659--708},
  year = {2000}
}

@article{EggersAnnuRevFluidMech2025,
  author = {Eggers, J. and Sprittles, J. E. and Snoeijer, J. H.},
  title = {Coalescence dynamics},
  journal = {Annual Review of Fluid Mechanics},
  volume = {57},
  pages = {61--87},
  year = {2025}
}

@article{EggersJFM1999,
  author = {Eggers, J. and Lister, J. R. and Stone, H. A.},
  title = {Coalescence of liquid drops},
  journal = {Journal of Fluid Mechanics},
  volume = {401},
  pages = {293--310},
  year = {1999}
}

@book{Faraday1839,
  author = {Faraday, M.},
  title = {Experimental Researches in Electricity, Volume I},
  publisher = {Richard and John Edward Taylor},
  address = {London},
  year = {1839}
}

@article{PachitariuNatMethods2022,
  author = {Pachitariu, M. and Stringer, C.},
  title = {Cellpose 2.0: how to train your own model},
  journal = {Nature Methods},
  volume = {19},
  pages = {1634--1641},
  year = {2022}
}

@article{StringerNatMethods2021,
  author = {Stringer, C. and Wang, T. and Michaelos, M. and Pachitariu, M.},
  title = {Cellpose: a generalist algorithm for cellular segmentation},
  journal = {Nature Methods},
  volume = {18},
  pages = {100--106},
  year = {2021}
}

@article{ThoroddsenPhysFluids2005,
  author = {Thoroddsen, S. T. and Etoh, T. G. and Takehara, K. and Ootsuka, N.},
  title = {On the coalescence speed of bubbles},
  journal = {Physics of Fluids},
  volume = {17},
  pages = {071703},
  year = {2005},
  doi = {10.1063/1.1965692}
}

@article{ChenAIChEJ2017,
  author = {Chen, Rou and Yu, Huidan W. and Zhu, Likun and Patil, Raveena M. and Lee, Taehun},
  title = {Spatial and temporal scaling of unequal microbubble coalescence},
  journal = {AIChE Journal},
  volume = {63},
  number = {4},
  pages = {1441--1450},
  year = {2017},
  doi = {10.1002/aic.15504}
}

@article{HossainPRE2022,
  author = {Hossain, Syed Sahil and Bashkatov, Aleksandr and Yang, Xuegeng and Mutschke, Gerd and Eckert, Kerstin},
  title = {Force balance of hydrogen bubbles growing and oscillating on a microelectrode},
  journal = {Physical Review E},
  volume = {106},
  number = {3},
  pages = {035105},
  year = {2022},
  doi = {10.1103/PhysRevE.106.035105}
}

@article{LuCEJ2024,
  author = {Lu, Xinlong and Yadav, Devendra and Zhou, Liwu and Li, Xiaoping and Ma, Lijing and Jing, Dengwei},
  title = {Evolution of hydrogen bubbles on a microelectrode driven by constant currents and its impact on potential response},
  journal = {Chemical Engineering Journal},
  volume = {500},
  pages = {156890},
  year = {2024},
  doi = {10.1016/j.cej.2024.156890}
}

@article{ZhangPRL2025,
  author = {Zhang, Yixin and Zhang, Xiangyu and Lohse, Detlef},
  title = {Why and When Merging Surface Nanobubbles Jump},
  journal = {Physical Review Letters},
  volume = {135},
  number = {19},
  pages = {194001},
  year = {2025},
  doi = {10.1103/s5pj-vmq8}
}

\end{document}